\documentclass[aps,prl,10pt,twocolumn,superscriptaddress]{revtex4-1}

\usepackage{times}
\usepackage{changes}

\usepackage{graphicx}
\usepackage{amssymb}
\usepackage{epstopdf}
\usepackage{amsmath}
\usepackage{color}
\usepackage{hyperref}
\usepackage{bbold}
\usepackage{enumitem}
\usepackage[acronym]{glossaries}
\usepackage{physics}
\makeatletter

\@ifundefined{textcolor}{}
{
	\definecolor{BLACK}{gray}{0}
	\definecolor{WHITE}{gray}{1}
	\definecolor{RED}{rgb}{1,0,0}
	\definecolor{GREEN}{rgb}{0,1,0}
	\definecolor{BLUE}{rgb}{0,0,1}
	\definecolor{CYAN}{cmyk}{1,0,0,0}
	\definecolor{MAGENTA}{cmyk}{0,1,0,0}
	\definecolor{YELLOW}{cmyk}{0,0,1,0}
}

\graphicspath{{Images/}}

\hypersetup{
	pdfpagemode={UseOutlines},
	bookmarksopen,
	pdfstartview={FitH},
	colorlinks,
	linkcolor={black},
	citecolor={blue},
	urlcolor={blue}}

\usepackage{hyperref}
\hypersetup{
	colorlinks=true,
	linkcolor=black,          
	citecolor=blue,           
	filecolor=magenta,        
	urlcolor=cyan
}

\makeatother	

\begin{document}
	
	\title{To heat or not to heat: time crystallinity and finite-size effects in clean Floquet systems}
	
	\author{Andrea Pizzi}
	\affiliation{Cavendish Laboratory, University of Cambridge, Cambridge CB3 0HE, United Kingdom}
	\author{Daniel Malz}
	\affiliation{Max-Planck-Institute of Quantum Optics, Hans-Kopfermann-Str. 1, 85748 Garching, Germany}
	\affiliation{Munich Center for Quantum Science and Technology (MCQST), 80799 Munich, Germany}
	\author{Giuseppe De Tomasi}
	\affiliation{Cavendish Laboratory, University of Cambridge, Cambridge CB3 0HE, United Kingdom}
	\affiliation{Max-Planck-Institut f{\"u}r Physik komplexer Systeme, N{\"o}thnitzer Stra{\ss}e 38, 01187-Dresden, Germany}
	\author{Johannes Knolle}
	\affiliation{Department of Physics, Technische Universit{\"a}t M{\"u}nchen, James-Franck-Stra{\ss}e 1, D-85748 Garching, Germany}
	\affiliation{Munich Center for Quantum Science and Technology (MCQST), 80799 Munich, Germany}
	\affiliation{Blackett Laboratory, Imperial College London, London SW7 2AZ, United Kingdom}
	\author{Andreas Nunnenkamp}
	\affiliation{School of Physics and Astronomy and Centre for the Mathematics and Theoretical Physics of Quantum Non-Equilibrium Systems, University of Nottingham, Nottingham, NG7 2RD, United Kingdom}
	
	\begin{abstract}
		A cornerstone assumption that most literature on discrete time crystals has relied on is that homogeneous Floquet systems generally heat to a featureless infinite temperature state, an expectation that motivated researchers in the field to mostly focus on many-body localized systems. Some works have however shown that the standard diagnostics for time crystallinity apply equally well to clean settings without disorder. This fact raises the question whether an homogeneous discrete time crystal is possible in which the originally expected heating is evaded. Studying both a localized and an homogeneous model with short-range interactions, we clarify this issue showing explicitly the key differences between the two cases. On the one hand, our careful scaling analysis confirms that, in the thermodynamic limit and in contrast to localized discrete time crystals, homogeneous systems indeed heat. On the other hand, we show that, thanks to a mechanism reminiscent of quantum scars, finite-size homogeneous systems can still exhibit very crisp signatures of time crystallinity. A subharmonic response can in fact persist over timescales that are much larger than those set by the integrability-breaking terms, with thermalization possibly occurring only at very large system sizes (e.g., of hundreds of spins). Beyond clarifying the emergence of heating in disorder-free systems, our work casts a spotlight on finite-size homogeneous systems as prime candidates for the experimental implementation of nontrivial out-of-equilibrium physics.
	\end{abstract}
	
	\maketitle
	
	\section{Introduction}
	In the past few years, a terrific amount of excitement has been raised around discrete time crystals (DTCs) \cite{wilczek2012quantum, sacha2015modeling, khemani2016phase, else2016floquet, yao2017discrete, Gambetta2019, yao2020classical, pizzi2019higher, pizzi2019period, pizzi2020bistability, smits2018observation, surace2018floquet, zhao2019, zhu2019dicke, malz2020seasonal, machado2020long, gong2018discrete, choi2017observation, zhang2017observation, rovny2018observation, else2017prethermal, else2020long, giergiel2018discrete, machado2020long, matus2019fractional, kshetrimayum2020stark}. In essence, DTCs are periodically driven systems characterized by a subharmonic response at a fraction of the drive frequency, thus breaking the discrete time-translational symmetry of the underlying equations \cite{sacha2015modeling, khemani2016phase, else2016floquet, yao2017discrete}. These nontrivial time phenomena are collective (or `many-body'), meaning that they crucially rely on the presence of infinitely many interacting elementary constituents, in complete analogy with, e.g., real-space crystals. In this sense, they extend the notion of quantum phase of matter to the non-equilibrium realm. According to the concept of universality \cite{Stanley1999}, the qualitative behavior of DTCs should rely, by definition, neither on the specific adopted model nor specific initial conditions. Rather, they should be robust to (weak) perturbations. At the same time, the breaking of time symmetry should not just be a transient phenomenon, but rather persist up to infinite time, analogous as to how order is maintained over arbitrary distances in a space crystal. In other words, a DTC should maintain an infinite autocorrelation time in spite of perturbations.
	
	Among the plethora of contexts in which time crystalline phenomena have been investigated, the original and arguably most studied one is that of a quantum spin chain with local interactions. The remainder of this work focuses on such a setting. In these systems, time crystallinity is typically proven by making use of exact diagonalization of finite-size systems, its main diagnostics being
	\begin{enumerate}[label=(\roman*)]
		\item the exponential scaling of the lifetime of the subharmonic response with system size \cite{else2016floquet, else2017prethermal, surace2018floquet} and
		\item the presence in the system's spectral response of a peak rigidly locked to a subharmonic frequency, not shifting under perturbations \cite{choi2017observation, pizzi2019higher, pizzi2019period, rovny2018observation, surace2018floquet, von2016absolute, yao2017discrete, zhang2017observation}.
	\end{enumerate}
	These two diagnostics are complimentary and interconnected. The fact that they make use of dynamical indicators is very natural, since the dynamics of local observables such as the magnetization $\langle\sigma^z_i \rangle (t)$ is accessible in experiments (in contrast, e.g., to the eigenstates).
	
	Under the unitary time evolution of a periodic (Floquet) Hamiltonian, short-range interacting systems are generally expected to thermalize, that is, to reach a featureless state at long times \cite{lazarides2014equilibrium}. This expectation shifted the focus to disordered models, which have the promise to evade the fate of thermalization through the mechanism of many-body localization (MBL) \cite{ponte2015periodically, ponte2015many, lazarides2015fate}. Indeed, it was shown that, under suitable conditions, all the eigenstates of the MBL Floquet operator come in pairs with quasienergy difference $\pi$, which underlines ergodicity breaking with a period-doubled subharmonic response for virtually any physically relevant initial condition: a sufficient condition for the realization of a DTC \cite{von2016absolute,von2016phase,moessner2017equilibration}.
	
	For these reasons, MBL systems legitimately gained a privileged position among the DTCs' candidates. The devil’s advocate, however, would argue that there is a priori no reason why the remarkable $\pi$-pairing condition on the eigenstates should be necessary for the observation of a DTC. This suspicion is motivated by the fact that physically relevant initial states such as ground states of local Hamiltonians or experimentally accessible states occupy only a small corner of the full Hilbert space. In the undriven setting, the lesson from quantum scars is in fact that, even in homogeneous many-body systems in which most of the eigenstates look completely ergodic, there might still be special initial conditions for which the dynamics resembles that of an integrable point, and thermalization is not straightforward \cite{turner2018weak}. Similarly, it has been argued that even homogeneous Floquet systems can, against the general expectation, evade thermalization \cite{prosen1998time, citro2015dynamical, haldar2018onset, pai2019dynamical}, as it was for instance shown for finite-size systems in the context of so-called dynamical localization \cite{ji2011nonthermal, ji2018suppression, fava2020many}. Therefore, if it is true that a natural definition of the DTCs is obtained through the dynamics of local variables, rather than at the level of the eigenstates, then it becomes legitimate not to take for granted that homogeneous Floquet systems necessarily thermalize, and to wonder whether DTCs may be possible in the absence of MBL. With this challenge in mind, the devil’s advocate would start testing the diagnostics (i) and (ii) on families of homogeneous Hamiltonians and initial conditions.
	
	Indeed, it turns out, this endeavour can be successful: the diagnostics (i) and (ii) on the duration and robustness of the subharmonic response, respectively, work equally well in the disordered and homogeneous scenarios, a fact that resulted in some papers claiming the existence of homogeneous (or ‘clean’) DTCs \cite{huang2018clean, yu2019discrete}. In these papers, time crystallinity is supported by analyses in complete analogy with those of the pioneering MBL DTCs, although convincing cases on \emph{why} the expected ergodicity should be broken are lacking. Very recently, a more compelling effort in trying to understand homogeneous DTCs has been made in Ref.~\cite{yarloo2020homogeneous}, where the authors ascribe time crystallinity to a scar-like mechanism preventing thermalization. The apparent success of the diagnostics (i) and (ii) for various homogeneous models appears confusing. The question whether MBL truly is necessary for a DTC is also posed by the observation of signatures of DTCs (using diagnostic (ii)) in experiments in which the role played by disorder and interaction range has remained unclear \cite{rovny2018observation, choi2017observation}.
	
	Here, we clarify this issue studying both an homogeneous and a MBL model. For finite-size systems $L < \infty$, we show that a robust and exponentially long-lived subharmonic response can emerge both in the presence and in the absence of localization, although its origin is different in the two cases. In MBL systems, this behavior is due to the well-known $\pi$-pairing mechanism involving \emph{all} the Floquet eigenstates, whereas in homogeneous systems it is due to the $\pi$-pairing of just two eigenstates. These have a large overlap with the initial condition in a way that resembles quantum scars \cite{turner2018weak}. We argue that special care is needed when adopting the above diagnostics (i) and (ii) to identify a DTC in a strict sense. In fact, it happens for the homogeneous case that not only the subharmonic response is exponentially long-lived (i) and robust (ii), but that its magnitude is also exponentially suppressed in system size, therefore disappearing in the thermodynamic limit $L \to \infty$, a fact that has been overlooked in the past.
	
	Our study leads to a twofold conclusion. On the one hand, we confirm that most likely no such a thing as a homogeneous DTC exists (in a strict sense and in a quantum short-range setting). On the other hand we show that crisp signatures of time crystallinity in these systems are nonetheless beyond doubt. In particular, the widespread belief that thermalization should occur over the timescale set by integrability-breaking terms \cite{else2020discrete} holds rigorously only in the thermodynamic limit, and very large finite-size systems can possibly behave nontrivially for much longer times. We show that under certain rather general and natural conditions thermalization only occurs at very large system sizes (of, e.g., many hundreds spins). We therefore argue that, although homogeneous DTCs may not exist according to the strict mathematical definitions, strong time crystalline signatures in large (but finite) homogeneous systems deserve much more consideration than they had in the past, as they are prime candidates for the observation of nontrivial dynamical phenomena.
	
	We emphasize that the phenomenology described here is markedly distinct from that of prethermal DTCs, which are characterized by subharmonic responses that are exponentially large in the driving frequency, rather than in system size \cite{abanin2017effective, abanin2017rigorous, else2017prethermal, machado2020long, pizzi2019higher, luitz2020prethermalization}. Indeed, in the thermodynamic limit we expect our system to behave in a thermal fashion, with heating happening over a timescale $\sim 1/V$, $V$ being the magnitude of the integrability breaking terms. The time crystalline signatures described here are due to remarkable finite-size effects and hold for very general (non-integrable) models.
	
	The remainder of this paper is organized as follows. First, we introduce the models. Second, we investigate the diagnostics (i) and (ii) and the subtleties of finite-size effects by means of careful scaling analyses. By inspecting the spectrum of the Floquet operator, we then reveal the mechanisms at the origin of the subharmonic responses in the two cases, and use a scaling analysis to explain why thermalization is ultimately expected in the thermodynamic limit. We conclude discussing the results and their implications for future research.
	
	\begin{figure*}[bth]
		\begin{center}
			\includegraphics[width=\linewidth]{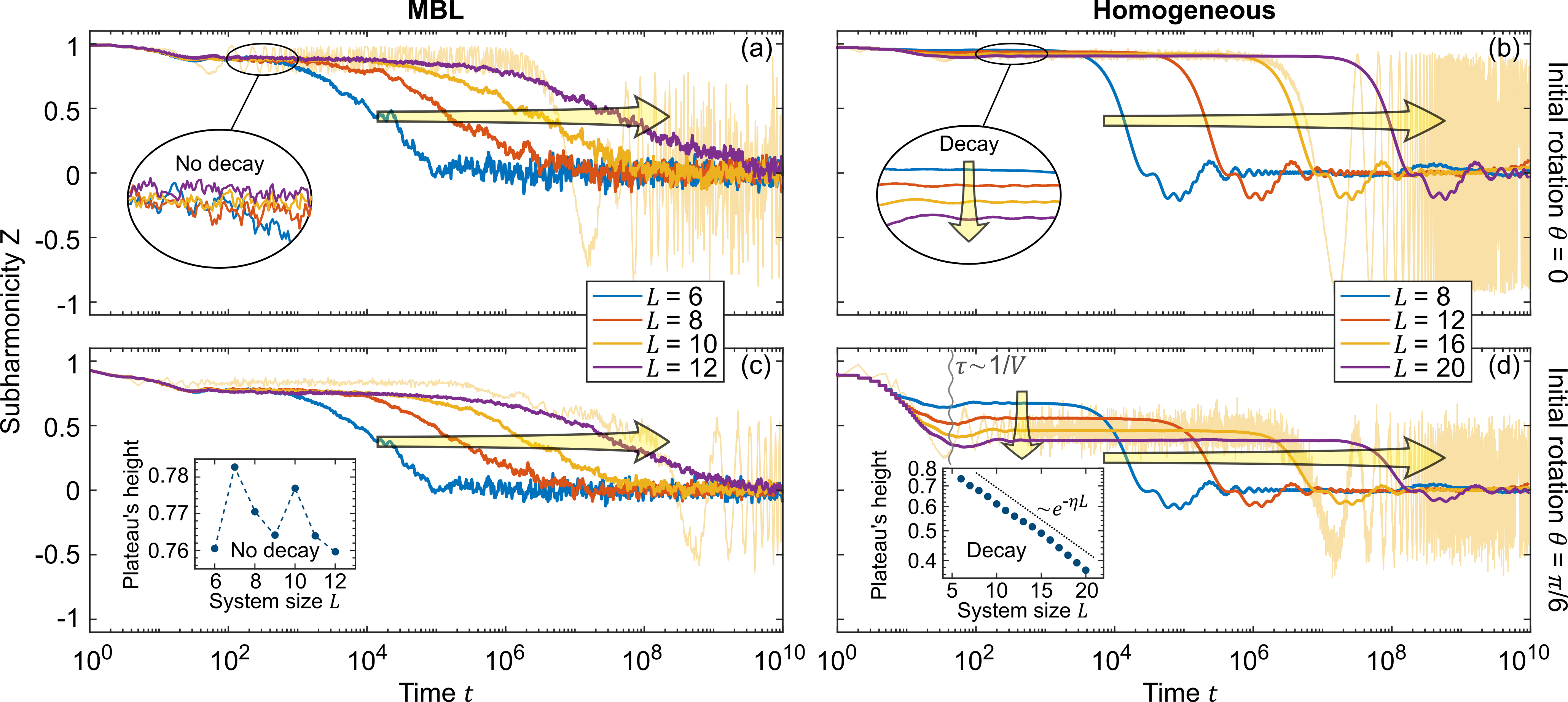}\\
		\end{center}
		\vskip -0.5cm \protect
		\caption{
			\textbf{Subharmonicity plateaus: a careful scaling analysis}. We investigate the scaling of the subharmonicity $Z(t)$ with system size, for both the MBL (a,c) and the homogeneous (b,d) models, and considering an initial rotation of the spins of angles $\theta = 0$ (a,b) and $\theta = \pi/6$. In all cases, the subharmonicity $Z(t)$ exhibits a crisp signature of time crystallinity: an exponentially long (in system size) plateau. In the MBL system, the pleateau's height does not decay with system size (insets), underpinning its stability and persistence in the thermodynamic limit $L \to \infty$. Contrary, in the homogeneous setting, the value of the plateau decays exponentially (insets), as can be better appreciated for larger $\theta$ (c,d), pointing towards the onset of thermalization in the thermodynamic limit, and to the impossibility of an homogeneous DTC in the strict sense. The value of the plateau's heights in (c,d) are obtained averaging $Z(t)$ over time. In (d), the plateau begins at a time $\tau \sim 1/V \sim 20$ (wiggly line), $V$ being the magnitude of the integrability breaking terms. For the MBL setting (a,c), results are averaged over 100 disorder realizations (one realization is plotted in faded colors for $L = 10$), whereas results in the homogeneous case (b,d) are averaged over a decade-long moving time window (one original time trace is plotted in faded colors for $L = 16$).}
		\label{FSfig1}
	\end{figure*}

	\section{Models}
	We consider a chain of $L$ interacting spins $1/2$ driven according to a binary protocol, as standard for DTCs. This protocol consists of the alternation of two Hamiltonians $H_1$ and $H_2$, resulting in the Floquet unitary operator
	\begin{equation}
	U_F = e^{-\frac{i H_2}{2}}e^{-\frac{i H_1}{2}},
	\end{equation}
	one cycle of the driving having period $T = 1$ ($\hbar = 1$). In the following, we focus on two very general models, one homogeneous and one MBL.
	As for the homogeneous model, we consider
	\begin{equation}
	H_{1} = \pi \sum_{j = 1}^L
	\sigma^x_j + \delta H_{1},
	\label{eq. H1c}
	\end{equation}
	\begin{equation}
	H_{2} = 2\sum_{j = 1}^L
 	\sigma^z_j \sigma^z_{j+1} + \delta H_{2},
	\label{eq. H2c}
	\end{equation}
	where $\sigma^x, \sigma^y$, and $\sigma^z$ denote the spin $1/2$ Pauli operators. The Hamiltonian $H_1$ describes an imperfect $\pi$-rotation, whereas the $ZZ$ coupling in $H_2$ should make the subharmonic response robust to perturbations.
	The terms $\delta H_{1,2}$ are small integrability breaking perturbations, and read
	\begin{equation}
	\delta H_{1,2} =
	2 \sum_{\nu = x, z} J^\nu_{1,2} \sum_{j = 1}^L \sigma^\nu_j \sigma^\nu_{j+1} +
	2 \sum_{\nu = x, z} h^\nu_{1,2} \sum_{j = 1}^L \sigma^\nu_j.
	\label{deltaHc}
	\end{equation}
	The parameter $h^x_1 = \pi/20$ describes the main imperfection in the $\pi$-rotation around the $x$ axis, $J^z_2 = 0$ without loss of generality, and the other coefficients are small ($\sim 0.05$) and, just to ward off any fine-tuning, integrability, or hidden symmetry, are drawn at random: $J^z_1 \approx 0.0306, J^x_1 \approx 0.0435, h^z_1 \approx 0.0134, J^x_2 \approx 0.0191, h^z_2 \approx 0.0546$ and $h^x_2 \approx 0.0550$.
	
	As for the localized model, we consider instead
	\begin{equation}
	H_{1} = \pi \sum_{j = 1}^L \sigma^x_j + \delta H_{1},
	\label{eq. H1MBL}
	\end{equation}
	\begin{equation}
	H_{2} = 2\sum_{j = 1}^L
	\left(J^z_{2,j} \sigma^z_j \sigma^z_{j+1} +
	h^z_{2,j} \sigma^z_j\right)
	+ \delta H_{2},
	\label{eq. H2MBL}
	\end{equation}
	where $J^z_{2,j}$ and $h^z_{2,j}$ are uniform random numbers in $[\frac{1}{2},\frac{3}{2}]$ and $[0,1]$, respectively, and where as before $\delta H_{1,2}$ are small integrability breaking perturbations
	\begin{equation}
	\delta H_{1,2} = 2\sum_{j = 1}^L \sum_{\nu = x, z}
	\left(J^\nu_{(1,2),j} \sigma^\nu_j \sigma^\nu_{j+1} +
	h^\nu_{(1,2),j} \sigma^\nu_j\right).
	\label{deltaHMBL}
	\end{equation}
	The parameter $h^x_{1,j} = \pi/40 + \delta h^x_{1,j}$ describes the main imperfection in the $\pi$-rotation, $J^z_{2,j} = h^z_{2,j} = 0$ without loss of generality, and all the remaining coefficients are uniform random numbers in $[0,0.01]$. Periodic boundary conditions are assumed for both models.
	
	As an initial condition, we take that considered by Else and collaborators in Ref.~\cite{else2016floquet}, that is
	\begin{equation}
	\ket{\psi_0} = e^{i \frac{\theta}{2} \sum_{j = 1}^{L} \sigma^x_j} \ket{\Uparrow},
	\label{eq. IC}
	\end{equation}
	where $\ket{\Uparrow}$ is the product state with all the spins polarized along $z$ and $\theta$ is the angle of rotation of the spins around the $x$ axis. Varying $\theta$, we can probe an entire family of physically relevant initial conditions. Note, our results are not contingent on the choice of Eq.~\eqref{eq. IC}, but rather hold for various families of experimentally relevant initial conditions, such as the ground states of some standard families of homogeneous Hamiltonians, see the Supplementary Material.
	
	Time crystallinity is investigated by means of two main observables.
	The first is used for the diagnostics (i) on the duration of the subharmonic response, may be called \emph{subharmonicity} \cite{pizzi2020bistability}, and is defined at stroboscopic times $t = 0,1,2,\dots$ as
	\begin{equation}
	Z(t) = (-1)^t \frac{1}{L} \sum_{j = 1}^{L} \langle \sigma_j^z (t) \rangle,
	\label{Eq. Z}
	\end{equation}
	where $\langle \dots \rangle$ denotes quantum expectation value and, in the MBL case, average over disorder realizations as well.
	In the presence of a period-doubled subharmonic dynamics, the spins rotate by an angle $\sim \pi$ at every Floquet period, $\langle \sigma_j^z (t) \rangle$ takes values $\sim +1, -1, +1, -1, \dots$ at times $t = 0, 1, 2, 3, \dots$, and $Z(t) \sim 1$. The parameter $Z(t)$ can be used to track the degree of subharmonicity of the response in time, and a finite and positive $Z(t)$ up to $t \to \infty$ is a signature of time crystallinity. By contrast, the relaxation of $Z(t)$ to $0$ corresponds to an ergodic behaviour.
	
	The second observable, useful for the diagnostics (ii) on the robustness of the subharmonic response, is the Fourier transform of the magnetization
	\begin{equation}
	\tilde{m}(f) = \frac{1}{M} \sum_{t = 0}^{M-1} e^{2\pi i f t} \frac{1}{L} \sum_{j = 1}^{L} \langle \sigma_j^z (t) \rangle,
	\label{Eq. mFFT}
	\end{equation}
	where $M$ is the number of Floquet periods over which the transform is computed. The presence in the spectral response $\tilde{m}$ of a peak at a subharmonic frequency $f = 0.5$ is a signature of time crystallinity, whereas, by contrast, no such a peak is found when the system behaves ergodically.
	
	\section{Results}
	Here, we present results obtained solving the models in Eqs.~(\ref{eq. H1c}, \ref{eq. H2c}) and Eqs.~(\ref{eq. H1MBL}, \ref{eq. H2MBL}) using exact diagonalization techniques.
	
	\textbf{Diagnostics for DTCs} --
	The subharmonicity $Z(t)$ in Eq.~\eqref{Eq. Z} is plotted in Fig.~\ref{FSfig1} to investigate the dynamics from the perspective of diagnostics (i). To start with, we consider a vanishing initial rotation $\theta = 0$ of the spins. Both in the MBL and homogeneous models, $Z(t)$ exhibits a non ergodic plateau of height $\sim 1$, whose length grows exponentially with system size, fulfilling the diagnostics (i) for time crstallinity. For this phenomenology, the standard argument would be that, in the thermodynamic limit $L \to \infty$, the period-doubled response is expected to extend up to infinite time, thus realizing a persistent DTC. Naively, one may think that this reasoning works equally well in the MBL as in the homogeneous case, but in the latter it is actually undermined by a subtle observation. After a more careful analysis, it turns out that the height of the plateau for the homogeneous model \emph{decreases} with system size, a crucial observation that, perhaps because hidden by time fluctuations, has not been reported before (to the best of our knowledge). In principle, the value of the plateau may therefore vanish in the thermodynamic limit, and the subharmonic response completely disappear.
	
	In the homogeneous setting, the decay of the plateau height is a warning sign, pointing towards thermalization in the thermodynamic limit. However, the range of plateaus values obtained for $\theta = 0$ is very limited ($\sim 5\%$), and an extrapolation to $L \to \infty$ is difficult. The decay of the plateau value is better appreciated for substantially larger perturbations of the initial condition, such as for $\theta = \pi/6$. In this case, the range of values ($\sim 65\%$) is broad enough to allow more confident conclusions on its scaling, that appear exponential. In striking contrast with the homogeneous scenario, the MBL model does not exhibit such a scaling (fluctuations of the plateau values are just due to noise, and are expected to disappear for a large enough number of disorder realizations). This confirms that, as expected, MBL systems can realize a robust DTC.
	
	In the thermodynamic limit, we expect the plateau to disappear and thermalization to occur over a timescale $\tau$  [marked with a wiggly line in Fig.~\ref{FSfig1}(d)]. We have checked that this timescale is set by the integrability breaking terms and scales as $\tau \sim 1/V$, $V$ being the magnitude of the intergability breaking terms in the Hamiltonian.
	
	\begin{figure}[bth]
		\begin{center}
			\includegraphics[width=1\linewidth]{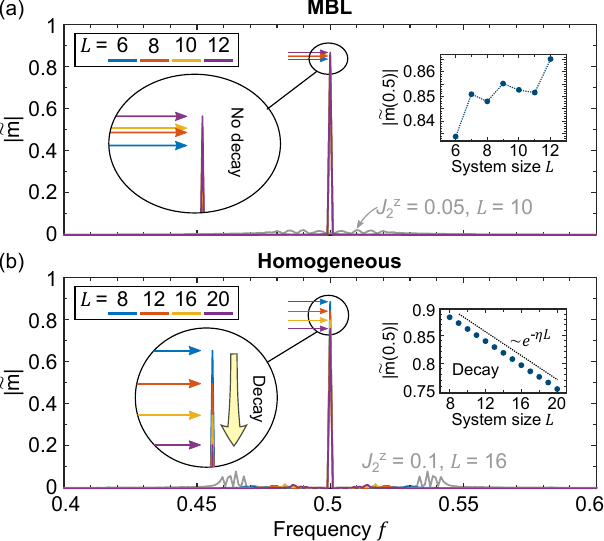}\\
		\end{center}
		\vskip -0.5cm \protect
		\caption{
			\textbf{Subharmonic peak: a careful scaling analysis.} For both the MBL (a) and the homogeneous (b) settings, we consider the diagnostics (ii) regarding the presence of a subharmonic peak in the system response. We plot the Fourier transform $\tilde{m}(f)$ of the magnetization computed over the first $10^3$ Floquet periods. In light grey, we report for reference the case of a small $J^z_2$, for which the subharmonic response disappears. The robust subharmonic responses are highlighted by a peak locked to the frequency $0.5$, in both cases. The difference between the homogeneous and the MBL scenarios is that the magnitude of the subharmonic peak does and does not decay with system size. Indeed, in the absence of MBL (b) we observe an exponential decay of $|\tilde{m}(0.5)|$ with system size, suggesting the disappearence of the subharmonic peak in the thermodynamic limit. Here, we considered an initial rotation of the spins $\theta = \pi/10$.}
		\label{FSfig2}
	\end{figure}

	\begin{figure*}[bth]
		\begin{center}
			\includegraphics[width=1\linewidth]{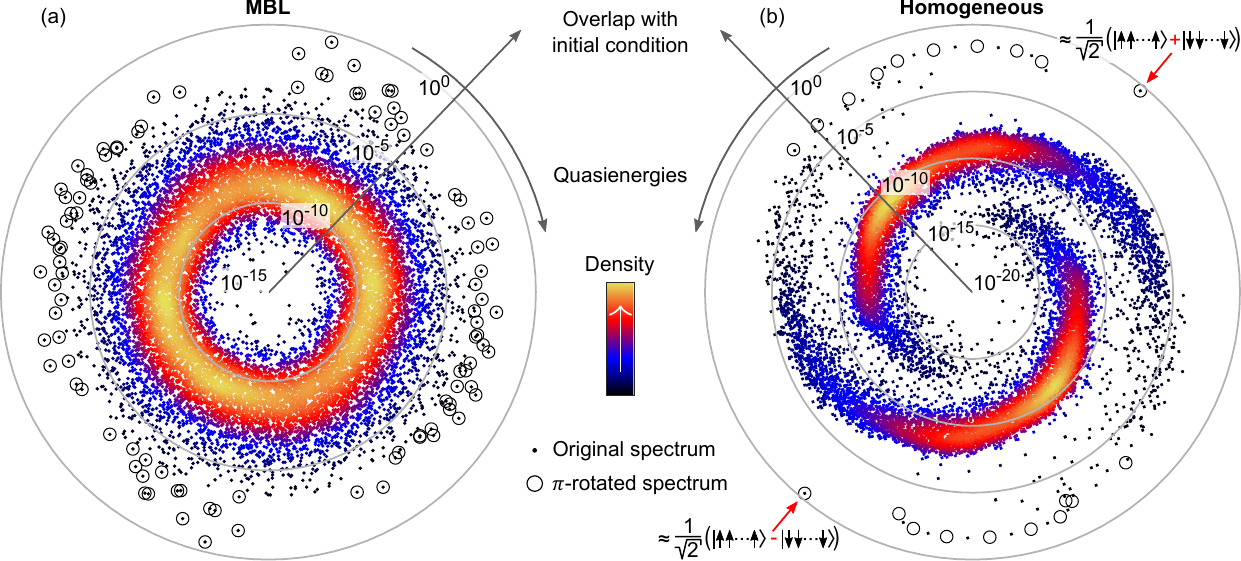}\\
		\end{center}
		\vskip -0.5cm \protect
		\caption{
			\textbf{Floquet spectra and the origin of subharmonicity.} Polar plots of the eigenstates of the Floquet operator $U_F$ for both the MBL ($L = 14$, a) and the homogeneous ($L = 19$, b) settings. The quasienergy and the overlap with the initial condition (for $\theta = \pi/10$) are used as angular and radial coordinates, respectively. The density of points is imprinted in the colorcode. Furthermore, some eigenstates are duplicated, rotated by a phase $\pi$, and plotted as circles. This way, two $\pi$-paired eigenstates are visually signalled by a dot centered in a circle. For graphical clarity, this duplication is only performed for the $100$ and $20$ eigenstates with the largest overlaps with the initial condition for the MBL and the homogeneous scenarios, respectively. (a) In the MBL case, all the eigenstates (or at least the considered outer ones) are $\pi$-paired, as expected. (b) In the homogeneous case, only the two outermost eigenstates (that is, those with largest overlap with the initial condition, highlighted with red arrows) are $\pi$-paired, whereas all the others are not. These two special eigenstates are approximately given by $\frac{1}{\sqrt{2}}\left(\ket{\Uparrow}\pm \ket{\Downarrow}\right)$, and their overlap with the initial condition determines the magnitude of the subharmonic response (e.g., of the plateaus' height in Fig.~\ref{FSfig1}, or of the subharmonic peak in Fig.~\ref{FSfig2}).}
		\label{FSfig3}
	\end{figure*}
	
	The system's spectral response, that is a standard probe for time crystallinity and the focus of diagnostics (ii), is instead investigated in Fig.~\ref{FSfig2}, where we plot the magnetization Fourier transform $\tilde{m}(f)$ in Eq.~\eqref{Eq. mFFT}. For both the MBL and the clean models, we verify the hallmark of time crystallinity: the presence of a peak rigidly locked to the subharmonic frequency $f = 0.5$ that does not shift in the presence of small perturbations. The genuine many-body nature of the phenomenon is observed for both cases, as the peak locks to the subharmonic frequency only for a large enough interaction. Again, subtle finite-size effects are appreciated by taking a close look at the scaling of the magnitude of the subharmonic peak: in the homogeneous (MBL) setting, the subharmonic peak decays exponentially (does not decay) in system size, which suggests its disappearence (persistence) in the thermodynamic limit, in complete analogy with Fig.~\ref{FSfig1}.
	
	\textbf{$\pi$-Pairing and the origin of the DTC behavior} -- 
	The results of Figs.~\ref{FSfig1} and \ref{FSfig2} show that the diagnostics (i) and (ii) work equally well in the MBL and in the homogeneous settings, but that in the latter clean case the subharmonic response is exponentially suppressed in system size. To gain a clearer intuition into this matter, we first have to better understand the origin of the subharmonic response in finite-size systems of the two types. To do so, in Fig.~\ref{FSfig3} we inspect the spectrum of the Floquet operator $U_F$. For a localized system, we confirm the expectation that \emph{all} the eigenstates come in pairs with quasienergy difference approaching $\pi$ exponentially in system size $L$. This $\pi$-pairing condition is distinctive of MBL DTCs, and is in striking constrast with non-localized systems. In the homogeneous setting, in fact, only two eigenstates are $\pi$-paired. Remarkably, these two special eigenstates are also those with the largest overlap with the considered initial condition, which explains why a subharmonic response is still observed, and why its intensity strongly depends on the initial rotation $\theta$. By inspection, we find that the two special eigenstates are approximately given by $\frac{1}{\sqrt{2}}\left(\ket{\Uparrow}\pm \ket{\Downarrow}\right)$, the approximation becoming an equality in the integrable limit (obtained for a drive with perfect $\pi$-flips and no perturbations, $J^{x,z}_{1,2} = h^{x,z}_{1,2} = 0$). We remark that, although the presence of a few non-thermal states may not be surprising in generic Hamiltonians at low energies, the $\pi$-pairing of two of them in the Floquet setting is.
	
\begin{figure}[bth]
		\begin{center}
			\includegraphics[width=1\linewidth]{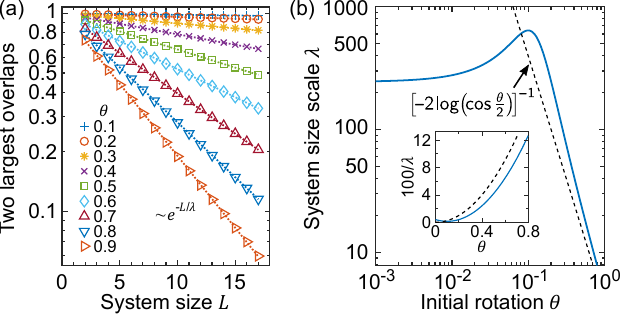}\\
		\end{center}
		\vskip -0.5cm \protect
		\caption{
			\textbf{Scaling of the two largest overlaps.} For the homogeneous setting, we investigate the dependence of the two largest overlaps, corresponding to the $\pi$-paired eigenstates, on the system size $L$ and initial rotation $\theta$. (a) For all considered values of $\theta$, the sum of the two overlaps decays as $e^{-L/\lambda}$ (exponential fits as dotted lines). (b) The characteristic system size scale $\lambda$ of the decay is plotted versus the initial rotation $\theta$. Even for small $\theta$, we find that $\lambda$ never diverges, meaning that thermalization is eventually expected in the thermodynamic limit. The scale $\lambda$ can take remarkably large values, up to a few hundreds, and is maximal for $\theta \approx 0.1$. The scaling in the integrable limit is plotted for reference as a dashed line.}
		\label{FSfig4}
	\end{figure}

	Since in the homogeneous case all the subharmonic response is ascribed to just two special eigenstates, it becomes crucial to study how their overlap with the initial condition scales with system size $L$. From the analysis in Fig.~\ref{FSfig4}, we find that such a scaling is clearly exponential $e^{-L/\lambda(\theta)}$, the decay occurring on a characteristic system size scale $\lambda(\theta)$ that depends on the initial rotation angle $\theta$. On the one hand, the finiteness of $\lambda < \infty$ suggests that, in the thermodynamic limit $L \to \infty$, the overlap with the two special states vanishes, and no subharmonic response survives at all, independent of initial condition. On the other hand, for $\theta \lessapprox 0.1$ we find that $\lambda$ takes very large values, in the order of a few hundreds. In this case, the onset of thermalization is appreciated only at remarkably large system sizes, way beyond the reach of exact and even approximate methods (such as density matrix renormalization group \cite{huang2018clean}). For instance, for the considered parameters, $\lambda$ has a maximum of $\approx 650 $ at $\theta \approx 0.1$, see Fig.~\ref{FSfig4}. An insight into the scaling is provided by the integrable limit, in which with a straightforward calculation one finds that the overlap approximately scales as $\left[\cos(\theta/2)\right]^{2L}$, that is, $\lambda(\theta) = \left[-2\log(\cos(\theta/2))\right]^{-1}$ (details in the Supplementary Material).

	\section{Discussion}
	In our analysis above, we showed that an exponentially long subharmonic response robust to perturbations can emerge in both the MBL and the homogeneous settings for finite-size systems. In the MBL case this behavior is due to a $\pi$-pairing mechanism involving all the eigenstates, whereas in the homogeneous setting it is instead due to the $\pi$-pairing of just two special eigenstates. This mechanism, which is genuinely many-body in nature, is reminiscent of quantum scars \cite{turner2018weak}, in which a few anomalous eigenstates that have a large overlap with the initial condition are responsible for weak ergodicity breaking after a quantum quench (that is, in the absence of a drive). There are however at least two important differences with respect to quantum scars in the non-driven setting. The first is that quantum scars are fragile to perturbations, whereas the $\pi$-pairing of the special eigenstates that we observe here is robust to perturbations of both the Hamiltonian and the initial condition, as long as these are homogeneous. The second is that the weak ergodicity breaking of quantum scars consists of a few oscillations before the ultimate onset of thermalization, occurring at relatively short timescales, whereas the subharmonic response here is exponentially long in system size. Furthermore, if the persistence of weak ergodicity breaking from quantum scars in the thermodynamic limit is still an open question \cite{surace2020exact}, the evidence suggests that in this limit the subharmonic dynamics in our model is completely suppressed.
	
	Indeed, our analysis suggests that the subharmonic response in homogeneous systems is a finite-size effect. The `critical system size' at which thermalization can be considered to take place was identified in the fitting parameter $\lambda$. It is worth noticing that the precise estimation of this critical size is a hard task, that may be undermined by even more severe finite-size effects. For instance, the analyses above were performed in a range of system sizes $L$ for which the spectrum of $H_2$ artificially splits into ergodic `minibands' (see Supplementary Section III), and one may expect that the scaling behavior could change abruptly at the critical $L_c$ at which the minibands of $H_2$ merge \cite{papic2015many}.
	
	Here highlighted in the Floquet scenario, subtle finite-size effects in many-body systems are known more generally to possibly occur in those many-body systems in which some length scales, such as the correlation and localization lengths, are larger than the system sizes amenable to exact techniques. Most prominently, in the non-driven setting, the debate around the existence and nature of the MBL phase has shown how finite-size effects can lead to controversial or misleading conclusions \cite{bera2017density, weiner2019slow, abanin2019distinguishing, sierant2020thouless}. A well-known example is that of the Anderson model on random-regular graphs \cite{altshuler2016nonergodic,tikhonov2016anderson}, for which the existence of a metal-insulator transition has been proven analytically \cite{abou1973selfconsistent}, and the value of its critical point is known within a few percents~\cite{parisi2019anderson, kravtsov2018non, tikhonov2019critical}. For this model, exact diagonalization on small systems points to an incorrect critical point, and a naive analysis could suggest the absence of the localized phase \cite{abanin2019distinguishing} and the existence of a highly debated critical/multifractal phase \cite{altshuler2016nonergodic,tikhonov2016anderson}. Our study draws the attention on analogous subtleties in the context of periodically-driven systems and DTCs.
	
	Finally, we remark that the surprising finite-size effects studied here are likely of broad applicability. Indeed, our spin model in Eqs.~(\ref{eq. H1c}, \ref{eq. H2c}) is general, and the observation of similar phenomenologies in previous studies on (non-integrable) systems of hard-core bosons \cite{huang2018clean} and spinless fermions \cite{yarloo2020homogeneous} suggests that analyses similar to ours may apply to quite generic homogeneous Floquet systems.
	
	\section{Conclusions}
	In conclusion, we clarified the issue whether MBL is truly needed to evade heating in a Floquet and short-ranged scenario. This issue has been ultimately raised by the fact that, as we observed, the standard diagnostics for DTCs (defined for finite-size systems) are fulfilled both in the MBL and in the homogeneous settings. We clarified it by observing that, on top of these diagnostics, there is the fact that in the clean scenario the subharmonic response is a finite-size effect, and its intensity (e.g., the magnitude of the subharmonic peak) decays exponentially in system size. This, to the best of our knowledge, has never been clarified before. Our results lead to the confirmation that only MBL systems can realize a DTC according to its strictest definition, stable to perturbations and persistent to infinite times in the thermodynamic limit.
	
	Nonetheless, what is remarkable is that the subharmonic response in homogeneous systems can be observed for many decades already for relatively small system sizes (e.g., $\sim 10$), whereas its weakening is possibly observed only at much larger sizes (e.g., $\sim 10^2$ or perhaps even $\sim 10^3$). This mismatch makes clean moderate-size systems a unique opportunity for implementation. Indeed, nowadays quantum simulators are typically limited to a few dozen elementary units \cite{arute2019quantum, ippoliti2020many}, and their coherence times are way below the timescales (e.g., of $10^{10}$ drive periods) that are considered in theoretical works such as ours. If the important and fundamental questions regarding the stability of a DTC in the limits $L \to \infty$ and $t \to \infty$ made MBL a necessity in theories dealing with strict mathematical definitions, this necessity is relaxed in most experimental scenarios, in which the remarkable exponentially long subharmonic response could be observed even in the absence of MBL. Moderate-size clean systems open therefore new avenues for the observation of time crystalline signatures in experiments, and for technological applications in the next generation of quantum devices.
	
	Finally, as a brief outlook for future research, it would be desirable to develop a general theory of finite-size clean Floquet matter able to capture the phenomenology and scalings observed here. As argued in the above Discussion, we expect our results to be of broad applicability, and an analytic theory could hopefully assess the generality of our conclusions on rigorous grounds. Moreover, further investigation should clarify how the transition between a strict MBL-DTC and finite-size DTC signatures relates to the MBL transition. This question could be addressed, for instance, by interpolating between the clean and MBL models in Eqs.~(\ref{eq. H1c}, \ref{eq. H2c}) and Eqs.~(\ref{eq. H1MBL}, \ref{eq. H2MBL}), respectively.
	
	\textbf{Acknowledgements.}
	We thank D.~Abanin for insightful comments on the manuscript. We acknowledge support from the Imperial-TUM flagship partnership. A.~P.~acknowledges support from the Royal Society. D.~M.~acknowledges funding from ERC Advanced Grant QENOCOBA under the EU Horizon 2020 program (Grant Agreement No.~742102). A.~N.~holds a University Research Fellowship from the Royal Society.

	\bibliography{DTCFiniteSize}

	\clearpage
	
	\setcounter{equation}{0}
	\setcounter{figure}{0}
	\setcounter{page}{1}
	\thispagestyle{empty} 
	\makeatletter 
	\renewcommand{\thefigure}{S\arabic{figure}}
	\renewcommand{\theequation}{S\arabic{equation}}
	\setlength\parindent{10pt}
	
	\onecolumngrid
	
	\begin{center}
		{\fontsize{12}{12}\selectfont
			\textbf{Supplementary Information for\\``To heat or not to heat: time crystallinity and finite-size effects in clean Floquet systems"\\[5mm]}}
		{\normalsize Andrea Pizzi, Daniel Malz, Giuseppe De Tomasi, Johannes Knolle, and Andreas Nunnenkamp \\[1mm]}
	\end{center}
	\normalsize

	These Supplementary Information are devoted to technical details and complimentary results, with focus only on the homogeneous setting of Eqs.~(2-4). More specifically, in Section I we compute the scaling constant $\lambda$ in the integrable limit, in Section II we consider a different family of initial conditions as compared to the main text, and in Section III we investigate the level statistics of the Hamiltonian $H_2$.
	
	\section{I) Integrable limit}
	Here, we study the integrable limit of Eqs.~(2-4), that is, we consider
	\begin{equation}
	H_{1} = \pi \sum_{j = 1}^L
	\sigma^x_j,
	\quad
	H_{2} = 2\sum_{j = 1}^L
	\sigma^z_j \sigma^z_{j+1},
	\label{eq. H12c int}
	\end{equation}
	with $h^x_{1} = \frac{\pi}{2}$ and $J^z_{2} = 1$. In this case, the Hamiltonian $H_1$ acts performing a perfect $\pi$-flip of the spins from $\ket{\uparrow}$ to $\ket{\downarrow}$ and viceversa, whereas $H_2$ has no effect beyond adding a phase. In this simple integrable limit, it is straightforward to see that the eigenstates of the Floquet operator are given by
	\begin{equation}
	\ket{s, \pm} = \frac{\ket{s} \pm \ket{\bar{s}}}{\sqrt{2}},
	\end{equation}
	where $\ket{s}$ is a product state of spins in the eigenstates $\ket{\uparrow}$ and $\ket{\downarrow}$ of the operators $\sigma_j^z$, and $\ket{\bar{s}}$ is its complimentary, with $\ket{\bar{\uparrow}} = \ket{\downarrow}$, and $\ket{\bar{\downarrow}} = \ket{\uparrow}$. It is simple to verfy that the states $\ket{s, +}$ and $\ket{s, \pm}$ have quasienergy difference $\pi$.
	
	When deviating from the integrable limit, the eigenstates and their quasienergies get perturbed, and the $\pi$-pairing condition is broken. As we have shown in the main text, there is nonetheless a pair of eigenstates whose quasienergies difference remains exponentially close to $\pi$. These eigenstates are those originating from
	\begin{equation}
	\ket{\Uparrow, \pm} = \frac{\ket{\Uparrow} \pm \ket{\Downarrow}}{\sqrt{2}},
	\end{equation}
	and it becomes therefore important to understand how these overlap with the initial condition. We compute this overlap with a straightforward calculation. We recall that the initial condition is given by
	\begin{equation}
	\ket{\psi_0} = e^{i \frac{\theta}{2} \sum_{j = 1}^{L} \sigma^x_j} \ket{\Uparrow}
	= \bigotimes_{j = 1}^L e^{i \frac{\theta}{2} \sigma^x_j} \ket{\uparrow}_j
	= \bigotimes_{j = 1}^L \left(\cos(\frac{\theta}{2})\ket{\uparrow} + i \sin (\frac{\theta}{2}) \ket{\downarrow}\right)_j.
	\end{equation}
	The overlap between $\ket{\psi_0}$ and $\ket{\Uparrow, +}$ is given by
	\begin{align}
		\left|\braket{\Uparrow,+}{\psi_0}\right|^2 = \frac{1}{2}\left|\braket{\Uparrow}{\psi_0} + \braket{\Downarrow}{\psi_0}\right|^2 +\dots 
		= \frac{1}{2}\left|\left[\cos(\frac{\theta}{2})\right]^L + \left[ i \sin (\frac{\theta}{2}) \right]^L\right|^2
		\approx \frac{1}{2} \left[\cos(\frac{\theta}{2})\right]^{2L},
	\end{align}
	where the last approximation holds for small enough $\theta$. The same result is obtained for the overlap of the initial condition with $\ket{\downarrow, -}$, and the sum of the two overlaps therefore reads
	\begin{equation}
	\left|\braket{\Uparrow,+}{\psi_0}\right|^2 + \left|\braket{\Uparrow,-}{\psi_0}\right|^2
	\approx \left[\cos(\frac{\theta}{2})\right]^{2L} = e^{-\eta L},
	\end{equation}
	with
	\begin{equation}
	\eta = \frac{1}{\lambda} = -2 \log(\cos(\frac{\theta}{2}))
	\end{equation}
	which is the result that we used in the main text.
	
	\section{II) Further evidences for the robustness to perturbations of the initial conditions}
\begin{figure}[bth]
		\begin{center}
			\includegraphics[width=1\linewidth]{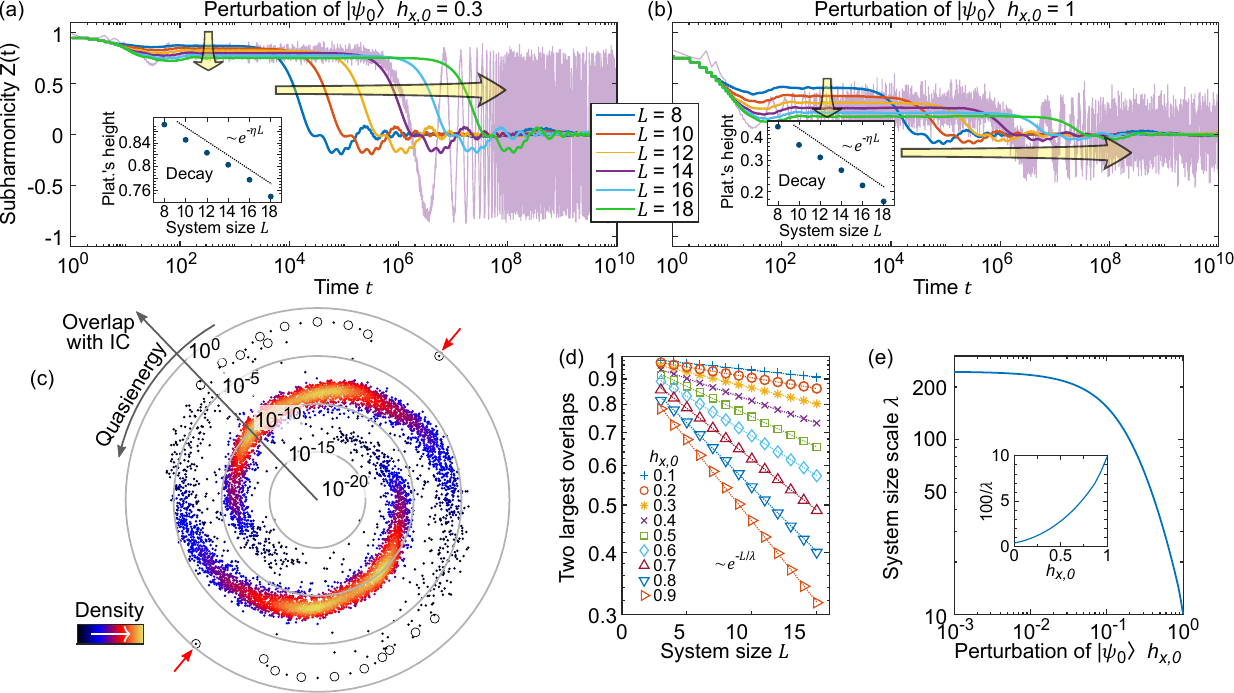}\\
		\end{center}
		\vskip -0.5cm \protect
		\caption{
			\textbf{Results for a different family of initial conditions.} We repeat some of the analyses of the main text considering as initial condition the ground state of the Hamiltonian in Eq.~\eqref{eq. H0}. The plots in (a), (b), (c), (d), and (e) are in complete analogy with Fig.~1(b), Fig.~1(d), Fig.~3(b), Fig.~4(a), and Fig.~4(b) in the main text, respectively. The only difference from the main text, to which we refer for a detailed interpretation of the results, is that the perturbation of the initial condition from the completely polarized state $\ket{\Uparrow}$ is here parametrized in $h_{x,0}$, rather than in $\theta$.}
		\label{FSfigS1}
	\end{figure}

	In the main text, we considered as initial condition the state that is obtained from the fully $z$-polarized state $\ket{\Uparrow}$ by applying a global rotation around the $x$-axis via the unitary $\exp(i \frac{\theta}{2} \sum_{j = 1}^{L} \sigma^x_j)$, in Eq.~(8). In this sense, the parameter $\theta$ played the role of magnitude of the perturbation of the initial condition. Here, we show that the results of our work are not contingent on this choice of initial conditions, but rather hold for perturbations of the state $\ket{\Uparrow}$ more in general, at least as long as these are still invariant under translations. In particular, we now investigate another kind of initial condition, that is the ground states of the Hamiltonian
	\begin{equation}
	H_0 = \sum_{j = 1}^{L} \left(\sigma_j^z \sigma_{j+1}^z + \frac{1}{10} \sigma_j^z + h_{x,0} \sigma_j^x \right).
	\label{eq. H0}
	\end{equation}
	For $h_{x,0} = 0$, the initial state is $\ket{\psi_0} = \ket{\Uparrow}$. Varying $h_{x,0} \neq 0$, the initial state $\ket{\psi_0}$ changes. The perturbation of the initial condition is therefore parametrized by the transverse field $h_{x,0}$. In Fig.~\ref{FSfigS1} we show that the very same analyses of the main text hold for this family of initial conditions. In particular, we show that the subharmonic response is exponentially long-lived, that it however decays with system size, that the speed of this decay is larger for larger perturbations $h_{x,0}$, that the phenomenology is due to the $\pi$-pairing of two special scarred eigenstates, and that the system size scale over which the subharmonic response is suppressed can be remarkably large.
	
	\section{III) Level statistics}
\begin{figure}[bt]
		\begin{center}
			\includegraphics[width=1\linewidth]{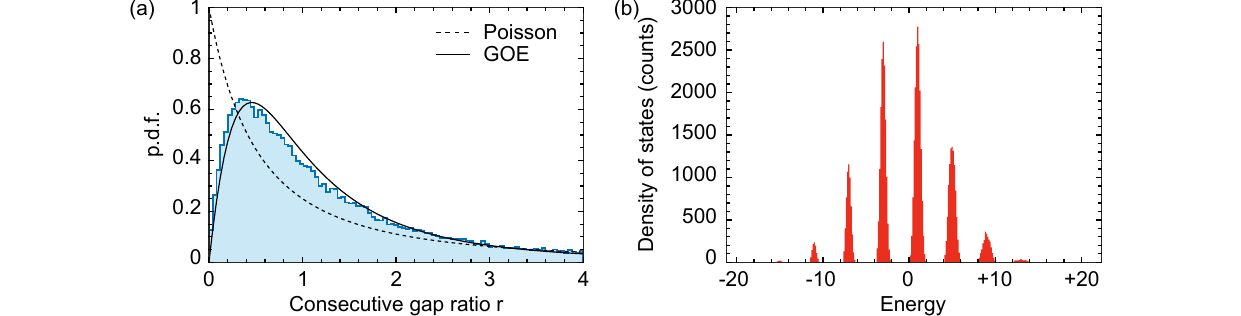}\\
		\end{center}
		\vskip -0.5cm \protect
		\caption{
			\textbf{Spectral properties of $H_2$.} For $L = 21$ sites and in the homogeneous settings, we investigate the spectrum of the Hamiltonian $H_2$. (a) The distribution of the consecutive level spacings ratio resembles that of a Gaussian orthogonal ensemble (GOE), highlighting the non-integrable nature of $H_2$. (b) Although non-integrable, the spectrum of $H_2$ is divided into separate bands, a qualitative feature likely due to finite-size effects.}
		\label{FSfigS2}
	\end{figure}

	The role of the perturbations in Eqs.~(2-4) is that to break integrability. To better understand to what extent integrability is broken, in Fig.~\ref{FSfigS2} we investigate the level statistics and the density of states of the spectrum of $H_2$ for the homogeneous setting. More specifically, we are interested in the distribution of the ratio between consecutive level spacings, that is
	\begin{equation}
	r_n = \frac{E_{n+1} - E_{n}}{E_{n} - E_{n-1}},
	\end{equation}
	with $E_n$ the eigenvalues of $H_2$ sorted in increasing order. The statistics of $r_n$ follows a GOE (Poisson) law with probability density function $P_{GOE}(r) = \frac{27}{8} \frac{r+r^2}{(1+r+r^2)^{5/2}}$ ($P_P(r) = \frac{1}{(1+r)^2}$) if the Hamiltonian is (is not) chaotic \cite{atas2013distribution}. On the one hand, the statistics of $r_n$ in Fig.~\ref{FSfigS2}(a) looks mostly chaotic. On the other hand, the energy levels in Fig.~\ref{FSfigS2}(b) are nonetheless organized in bands, whose separation may be a finite-size effect. In non-driven dynamical scenarios (e.g., quantum quenches), it has been argued that the presence of these bands makes finite-size effects particularly subtle and misleading \cite{papic2015many}, and it might be that the same extends to the driven setting in which $H_2$ is alternated with $H_1$, as considered in the main text. Indeed, one expects that the width of these bands increases with system size quicker than their separation, so that the bands touch at a critical system size $L_c$ (beyond the reach of exact diagonalization techniques). This touching may favour thermalization, and it is possible that the trends observed in the scaling analyses of the main text may change abruptly at $L = L_c$.
\end{document}